# Smart-phone Sensor of Pendulum Motion

Randall D. Peters

Physics Department
Mercer University, Macon, GA

**Abstract**

Described is an experiment where the embedded accelerometer of a smart-phone was used to study the free decay of a `simple' pendulum to which the phone was attached.

**Background**

An accelerometer is part of the internal hardware of the Droid-x smart-phone that was used in this study [1]. To control this accelerometer, a `seismo app' [2] was downloaded and installed in the operating system of the phone.

**Pendulum**

The pendulum to which the phone was attached for this study was fabricated as follows. Two small loops, one at each end, were formed on a commonplace type leather belt whose unmodified length is approximately 42 in. The Droid-x was placed in the bottom end-loop, and the vertically suspended belt was supported by a screwdriver placed in the top loop. At equilibrium of the pendulum, the x-axis of the accelerometer was nearly vertical (determined by the phone's orientation in the belt-loop). The handle of the horizontal screwdriver was placed on top of a file cabinet and held motionless, so that the pendulum would swing, without touching, in a plane that was parallel-to and close-to a side face of the cabinet.

For data acquisition, the accelerometer program was initialized, and the pendulum subsequently released from rest, with an initial displacement of about 45 degrees. The accelerations recorded during one free decay are shown in Fig. 1. Consistent with its near-vertical-at-equilibrium orientation, the largest component of acceleration is seen from the figure to be that of the x-axis.

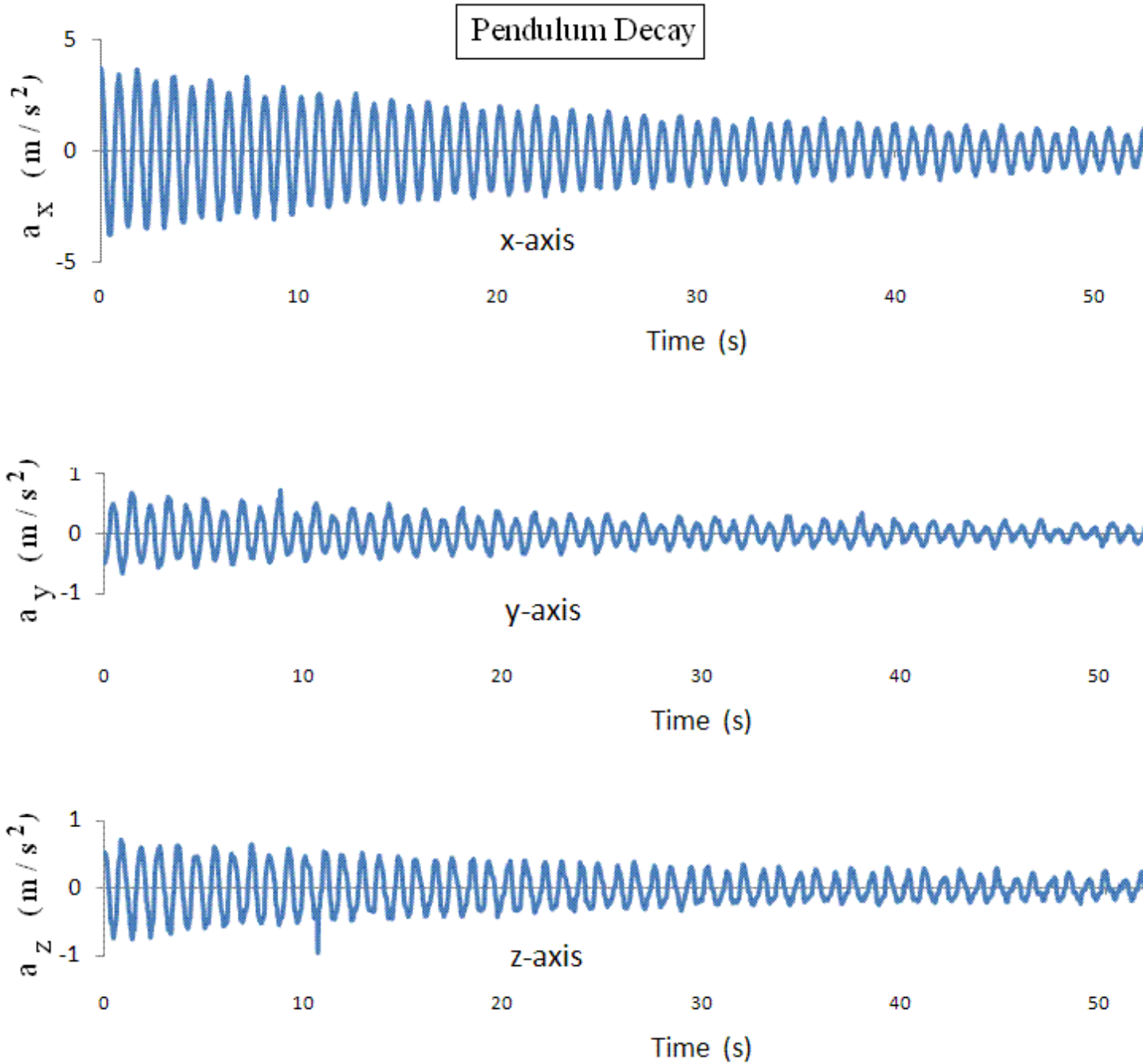

**Figure 1.** Time dependent accelerations recorded during the pendulum's free decay. A constant offset of 9.8 m/s$^2$ (earth's field) is not visible in the x-axis case, because the 'filter' function of the `seismo' algorithm was set to be operative during data acquisition.

All graphs of this article were generated using Microsoft Excel. A given record saved to Droid-x memory was transferred to the personal computer holding Excel, by means of e-mail (Google's gmail) attachment. This is the only means with which the author was able to gain access to the records.

Shown in Fig. 2 are frequency domain plots for each of the x, y, and z acceleration components. These were calculated with the Excel Fourier Transform algorithm (named `Fourier Analysis' by Microsoft). At the (fixed for `Seismo') sample rate of 30 Hz, each 52 s record contains 1560 samples. Since the FFT [3] operates only on records of length N = $2^n$ with n being an integer, only the first 34.13 s (1024 points, n = 10) were used in calculating a given spectrum.

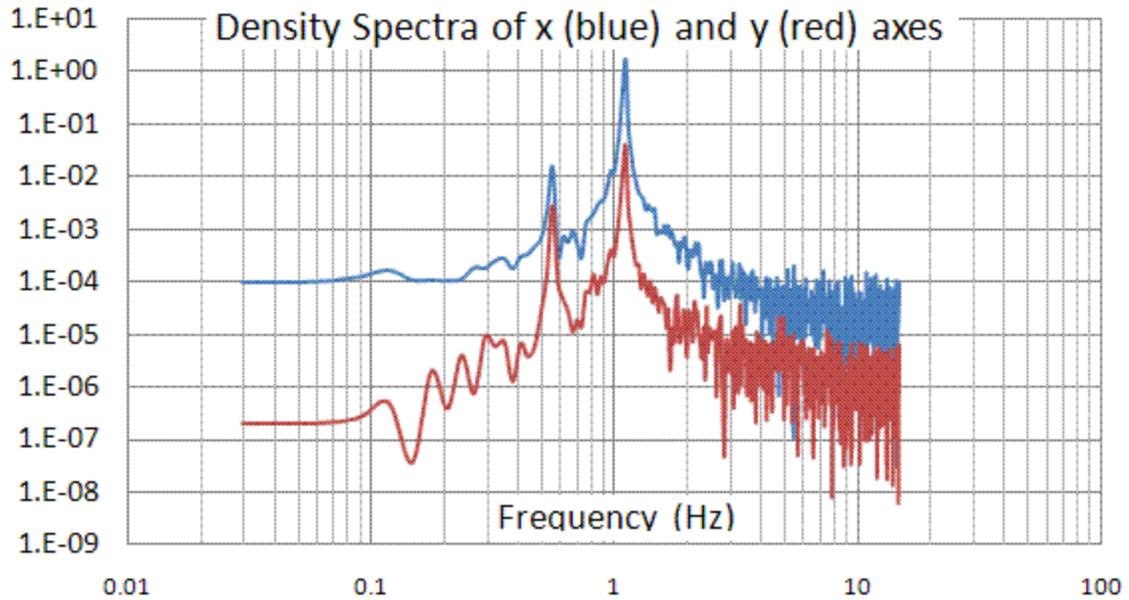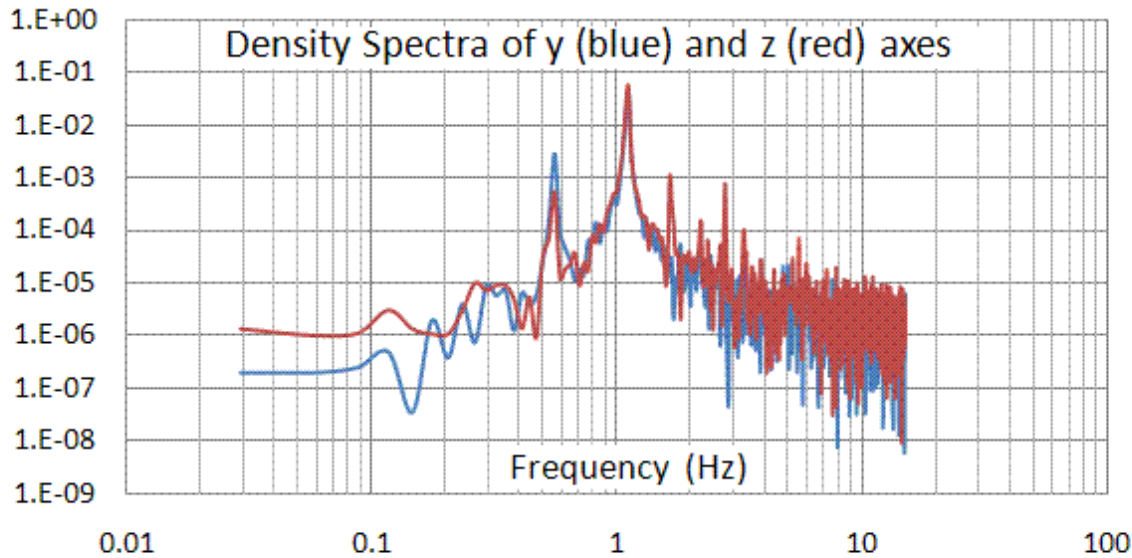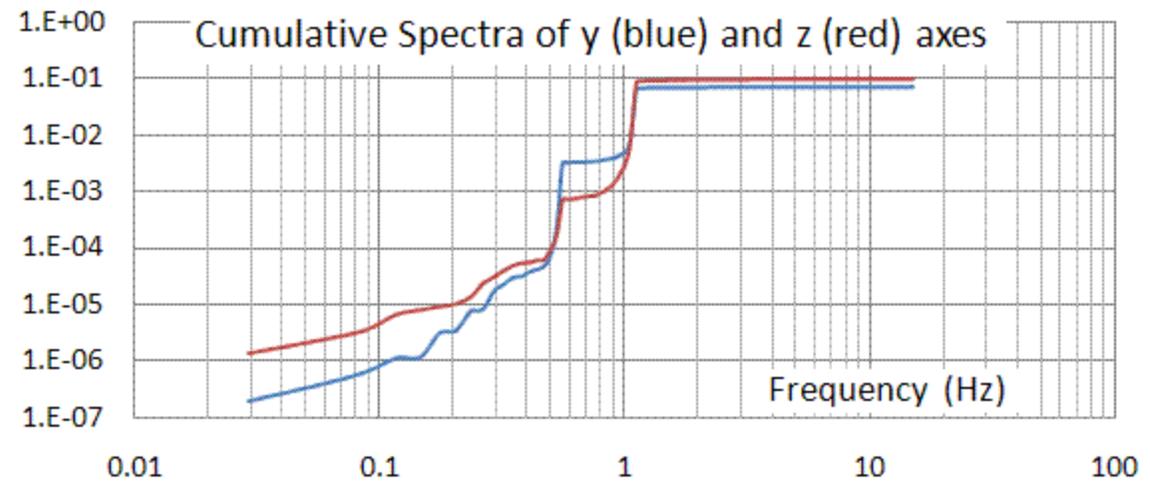

**Figure 2.** Frequency-domain plots for each of the three cases shown in Fig. 1. Only for components x and y (or z) is their magnitude difference large enough for the pair to be meaningfully plotted on the same density spectrum graph (such as the top plot).

In all of the frequency-domain graphs, the total number of points per plot is 512; i.e., one-half the number of points in the total transform-being thus the `single-sided' form that ignores negative frequency components of the FFT. Each ordinate value corresponds to the square of the FFT modulus multiplied by the factor of 2. Only index values in the range 1 to 512 were considered in the calculation, corresponding to the positive frequency components. It should be noted that the Excel FFT algorithm is (like most with which the author is familiar) one that does not yield normalized output. Normalization, so that Parseval's theorem is satisfied, requires that the Excel generated modulus values be each divided by $N^2$. Thus for the present work, each raw-modulus value was divided by $1024^2 = 1.049 \times 10^6$.

Unlike the considerable difference in magnitude between x and y components, those involving y and z are close to the same size. When superposed, their density spectrum graph is cluttered and hard to interpret (middle plot). This shortcoming is readily overcome by working instead with the cumulative spectrum, as shown in the lowest plot. The cumulative spectrum is obtained from the density spectrum by integrating over frequency from its lowest value upwards [4].

The primary mode of the pendulum at 1.11 Hz is clearly visible in every case. A lower frequency mode at 0.557 Hz is also visible in the plots, due to a periodic twisting of the leather belt. Unique to the z-axis are some higher frequency components that derive from flexures of the leather belt about an axis perpendicular to its width, which is much greater than its thickness. To readily see that these are present in the z-axis but not the y-axis, it is convenient to work with an alternative cumulative spectrum expressed in terms of period rather than frequency, as shown in Fig. 3.

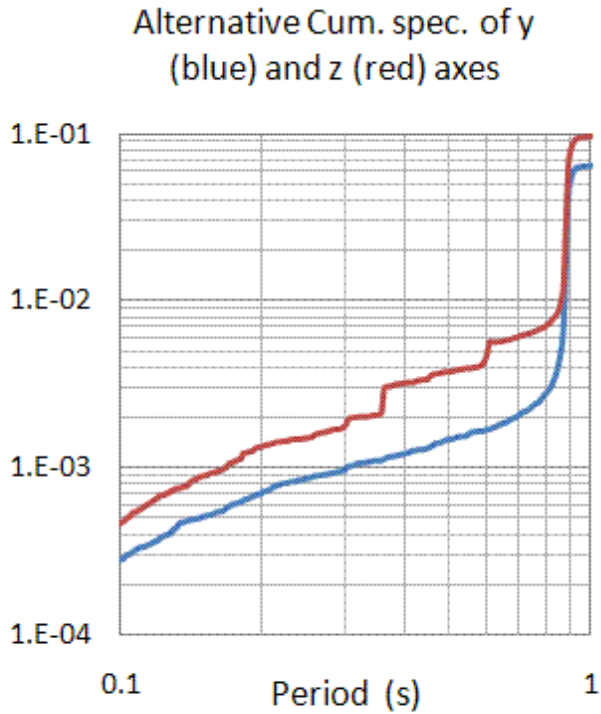

**Figure 3.** Period-domain plot showing two prominent high frequency modes associated only with the z-axis.

To calculate the alternative cumulative spectrum, the integrals are done in an opposite direction to that of the earlier plots; i.e., now they are from the highest frequency downward, even though the abscissa of the plot is in terms of period (reciprocal of frequency).